\documentclass[conference]{IEEEtran}
\IEEEoverridecommandlockouts
\usepackage{cite}
\usepackage{amsmath,amssymb,amsfonts}
\usepackage{bbm} 
\usepackage{algorithmic}
\usepackage{booktabs}
\usepackage[draft]{hyperref}
\usepackage{graphicx}
\usepackage{textcomp}
\usepackage{siunitx}
\usepackage{subcaption}
\usepackage{xcolor}
\usepackage{algorithm}
\usepackage{float}
\usepackage{stfloats}
\usepackage[short]{optidef}
\def\BibTeX{{\rm B\kern-.05em{\sc i\kern-.025em b}\kern-.08em
    T\kern-.1667em\lower.7ex\hbox{E}\kern-.125emX}}
\begin{document}

\title{Propagation and Rate-Aware Cell Switching Optimization in HAPS-Assisted Wireless Networks\\

}

\author{
\IEEEauthorblockN{
Mehmet Eren Uluçınar\IEEEauthorrefmark{1},
Özgün Ersoy\IEEEauthorrefmark{1},
Berk Ciloglu\IEEEauthorrefmark{2},
Metin Ozturk\IEEEauthorrefmark{1},
Ali Gorcin\IEEEauthorrefmark{3,}\IEEEauthorrefmark{4}}

\IEEEauthorblockA{\IEEEauthorrefmark{1}Electrical and Electronics Engineering, Ankara Yıldırım Beyazıt University, Ankara, Türkiye}
\IEEEauthorblockA{\IEEEauthorrefmark{2}Department of Information Engineering, University of Brescia, Brescia, Italy}
\IEEEauthorblockA{\IEEEauthorrefmark{3}\href{https://hisar.bilgem.tubitak.gov.tr/en/}{Communications and Signal Processing Research (HİSAR) Lab., TÜBİTAK BİLGEM, Kocaeli, Türkiye}}
\IEEEauthorblockA{\IEEEauthorrefmark{4}Department of Electronics and Communication Engineering, Istanbul Technical University, İstanbul, Türkiye}
}

\maketitle

\begin{abstract}
Cell switching is a promising approach for improving energy efficiency in wireless networks; however, existing studies largely rely on simplified models and energy-centric formulations that overlook key performance-limiting factors. This paper revisits the cell switching concept by redefining its modeling assumptions and mathematical formulation, explicitly incorporating realistic propagation effects such as building entry loss (BEL) and atmospheric losses relevant to non-terrestrial networks (NTN), particularly high-altitude platform station (HAPS). Beyond proposing a new cell switching strategy, the conventional energy-focused problem is reformulated as a multi-objective optimization framework that jointly minimizes power consumption, unconnected users, and data rate degradation. Through this reformulation, the proposed methods ensure that energy-efficient operation is achieved without compromising user connectivity and data rate performance, thereby inherently supporting sustainability objectives for sixth-generation (6G) networks. To solve this reformulated problem, two complementary approaches are employed: the weighted sum method (WSM), which enables flexible and adaptive weighting mechanism, and the $\epsilon$-constraint-inspired method ($\epsilon$CM), which converts connectivity and rate-related objectives into constraints within the conventional energy-focused problem. Moreover, unlike prior work relying only on simulations, this study combines system-level simulations with Sionna–OpenAirInterface (OAI) based emulation on a smaller network to validate the proposed cell switching concept under realistic conditions. The results show that, compared to the conventional approach, WSM reduces rate degradation for up to 70\% for high-loss indoor users and eliminates the 44\% drop for low-loss indoor users.
\end{abstract}


\section{Introduction}
The rapid growth in connected devices, such as Internet of things (IoT) applications, together with increasing data rate demands, has led to a significant rise in mobile traffic~\cite{ITUStat}. To address this trend, sixth-generation (6G) cellular communication networks are expected to enable features such as immersive communication and ubiquitous connectivity, and ultra-reliable low-latency communication~\cite{ITU}. However, existing radio access network (RAN) deployment scenarios are insufficient for these promising features to be realized, therefore, network densification through the deployment of additional base stations (BSs) has become a key approach to improving coverage, capacity, and data rates. This densification, however, significantly increases network energy consumption and directly conflicts with the sustainability goals of 6G~\cite{1}. More broadly, sustainability in 6G encompasses environmental, economic, and social dimensions, where energy-efficient network operation constitutes a key aspect of environmental sustainability, as energy consumption is a significant contributor to environmental impact in wireless networks~\cite{7}.

One common solution to reduce energy consumption and enable sustainable operation in 6G networks is the cell switching technique, where BSs with low or no load are put into sleep mode (i.e., switched off) to save energy. To offload the traffic of the switched-off BSs, non-terrestrial networks (NTN), particularly high-altitude platform station (HAPS), offer significant advantages through their enormous coverage footprint and ability to provide additional capacity~\cite{3}.

Nonetheless, there are a limited number of studies in the literature that integrate HAPS with cell switching. In~\cite{7}, HAPS is used as an offloading entity to manage sudden traffic surges in the terrestrial networks (TN), while \cite{8} proposes a hybrid aerial–terrestrial network combining HAPS and terrestrial BSs, where traffic from lightly loaded BSs is offloaded to HAPS. 
In~\cite{9}, the load factors of the SBSs in the network are sorted in ascending order, and the SBSs are sequentially switched off.
Furthermore, a HAPS–SBS network is proposed in ~\cite{10}, in which  the traffic of the switched-off SBSs is offloaded to HAPS. These studies demonstrate that HAPS has great potential to handle offloaded traffic and to reduce network energy consumption through cell switching. Nonetheless, most cell switching studies focus only on minimizing network power consumption, treating each user as a traffic load regardless of their environmental and propagation-related conditions. Incorporating situational awareness is therefore essential for a more robust cell switching. After cell switching, the BS providing the strongest received signal strength may be switched off, causing the user to be offloaded to a less optimal BS in terms of received signal strength. Specifically, atmospheric losses, arising from offloading users to HAPS, and higher BEL, particularly for indoor users, degrade received signal strength. Consequently, connectivity and data rate, being functions of signal strength, are both negatively affected after cell switching. Moreover, the literature includes only a limited number of studies that validate cell switching concepts through realistic or practical implementations.

In this study, we analyze the impacts of environmental variables on signal propagation loss and their effects on cell-switching performance. We capture the environmental propagation dynamics with building entry loss (BEL), which occurs when radio waves propagate through various types of materials. In addition, atmospheric losses, including gas and rain attenuation, are also considered, as they significantly affect user data rates. To account for these realistic propagation effects in cell-switching decisions, we reformulate the conventional energy-focused cell switching problem as a multi-objective optimization problem, and develop two complementary solution approaches:  the weighted sum method
(WSM) and the $\epsilon$-constrained–inspired method ($\epsilon$CM). This reformulation allows cell switching decisions to consider not only energy efficiency but also user connectivity and data rate, thereby addressing key practical limitations overlooked in previous studies. The main contributions of this work are as follows:

\begin{itemize}
    \item The impacts of various propagation losses, including BEL and atmospheric losses, on the cell switching performance are mathematically derived.
    \item The conventional energy-focused cell switching problem is reformulated as a multi-objective optimization problem by incorporating the number of connected users and their achievable rate objectives alongside power minimization. This reformulation addresses practical limitations of prior works, where neglecting user connectivity and rate degradation leads to solutions that are not readily applicable in real deployments.
    \item Two complementary approaches are developed to solve the multi-objective problem: (i) WSM combines the power, connectivity, and achievable rate objectives into a single objective function through a flexible and adaptive weighting mechanism that enables objective prioritization based on varying network requirements. (ii) $\epsilon$CM converts the connectivity and data rate objectives into constraints within the conventional energy-focused cell-switching, ensuring that sustainability gains are achieved without compromising user connectivity and data rate.
    \item A hybrid framework combining large-scale simulations with a Sionna–OpenAirInterface (OAI) emulation demonstrates the practical feasibility and real-world applicability of the proposed approach, a perspective rarely addressed in existing literature.
\end{itemize}

To the best of the authors’ knowledge, this work represents the first attempt to analyze various propagation parameters, such as BEL and atmospheric losses, within the context of the HAPS-assisted cell switching problem. In addition, reformulating the conventional cell switching problem as a multi-objective optimization using two complementary approaches constitutes another novel contribution to the literature.

\section{System Model}
\label{sec:System Model}

\subsection{Network Model}

In this work, a network model consisting of one HAPS equipped with a super-macro BS (HAPS-SMBS), one MBS, and multiple SBSs is considered. There are $\Gamma$ SBSs, indexed by \( s \in \mathcal{S} \), where $\mathcal{S} = \{1, 2, \ldots, \Gamma\}$. and $\Gamma +2$ total BSs, indexed by $b\in \mathcal{B}$, where $\mathcal{B}=\{1,2,...,\Gamma+2\}$. Furthermore, the systems serves $\chi$ UEs indexed by \( u \in \mathcal{U} \), where \quad $\mathcal{U} = \{1, 2, \ldots, \chi\}$. These users are distributed randomly throughout the network environment, which covers an area of 1 km-by-1 km. These users are divided into three categories with approximately equal cardinality (i.e., $\chi/3$); namely: high-loss ($\psi$), low-loss ($\kappa$), and outdoor users ($\rho$). Moreover, user mobility is modeled as a random walk, where users move in random directions with a fixed step size. 

\subsection{Propagation Model}

Since there are two types of networks (i.e., TN and NTN), the path loss is calculated separately.

\subsubsection{Path Loss Model for TN} 
In this study, the path loss between a TN BS and a UE is modeled according to the 3rd Generation Partnership Project (3GPP) Urban Macro (UMa) specifications \cite{12}, which incorporates both line-of-sight (LoS) and non-LoS (NLoS) conditions. The path loss for the LoS scenario in TN is computed by

\begin{equation}
\begin{split}
L_{\mathrm{LoS}} = 28 + 22 \log_{10}(d_{\mathrm{3D}}) + 20 \log_{10}(f_{\mathrm{GHz}}) \\ +  \sigma_{\mathrm{LoS}} + L_{\mathrm{BEL}} +  L_{\mathrm{E}} + \chi_{\mathrm{LoS}},    
\end{split}
\label{loss}
\end{equation}
where $d_{\mathrm{3D}}$ represents the three-dimensional distance between the BS and UE, and $f_{\mathrm{GHz}}$, $\sigma_{\mathrm{LoS}}$, $L_{\mathrm{BEL}}$ and $\chi_{\mathrm{LoS}}$ represent the carrier frequency, shadowing, BEL and small scale fading for LoS, respectively. Also, $L_\mathrm{E}$ denotes the additional loss associated with the user type \cite{ITU_R_P2346_3}. In a similar vein, the path loss for the NLoS scenario in NTN is evaluated by
\begin{equation}
\begin{split}
L_{{\mathrm{NLoS}}} = 32.54 + 30 \log_{10}(d_{\mathrm{3D}}) + 20 \log_{10}(f_{{\mathrm{GHz}}})\\
+ \sigma_{\mathrm{NLoS}} + L_{\mathrm{BEL}} + L_{\mathrm{E}} + \chi_{\mathrm{NLoS}},
\end{split}
\label{Nloss}
\end{equation}
where $\chi_{\mathrm{NLoS}}$ denotes the small scale fading for NLoS.

\subsubsection{Path Loss Model for NTN}
 The path loss model for the link between HAPS-SMBS and UE is obtained from the 3GPP report in \cite{12}. The signal undergoes various path losses between the HAPS-SMBS and the UE, and $L_{\mathrm{HAPS}}$, the total path loss, can be obtained by

\begin{equation}
L_{\mathrm{HAPS-SMBS}} = L^{}_{\mathrm{FSPL}} + L_{\mathrm{a}} + L_{\mathrm{BEL}} + L_{\mathrm{E}},
\label{Hloss}
\end{equation}
where $L_{\mathrm{FSPL}}$ and $L_{\mathrm{a}}$ are the free space path loss (FSPL) and attenuation caused by atmospheric losses, respectively. $L_{a}$ includes gas and rain attenuations.

\subsection{Power Consumption Model}
The EARTH power model, as introduced in~\cite{13}, is adopted in this work to calculate the instantaneous power consumption  of BSs. Thus, the power consumption $P_{i,t}$, of BS $i$ at time $t$ is calculated as

\begin{equation}
P_{i,t} =
\begin{cases}
P_{\mathrm{O},i} + \eta_i \lambda_{i,t} P_{\mathrm{T},i}, & 0 < \lambda_{i,t} < 1, \\[6pt]
P_{\mathrm{S},i}, & \lambda_{i,t} = 0,
\end{cases}
\label{power}
\end{equation}
where $P_{\mathrm{T},i}$ is the BS transmit power, $P_{\mathrm{O},i}$ is the operational circuit power, $\eta_i$ is the efficiency, $P_{\mathrm{S},i}$ is the sleep circuit power, and $\lambda_{i,t}$ is the instantaneous load factor at time $t$.
\subsection{User Allocation}

User allocation is implemented based on the SINR received by each user from each BS. The SINR received by user $u$ from BS $b$ is obtained by

\begin{equation}
\gamma_{u,b} = 
\frac{P_{u,b}}
{P_{\mathrm{n}} + \displaystyle\sum_{k \in \mathcal{B}_{\mathrm{act}} \setminus \{b\}} P_{u,k}},
\label{SINR}
\end{equation} 
where $P_{u,b}$, $P_{\mathrm{n}}$, and $P_{u,k}$ denote the received power from BS $b$, the noise power, and interference power, respectively.

To assign users to a BS, three conditions must be satisfied: 

\emph{i)} The BS should provide the highest SINR value compared to other BSs, such that
\begin{equation}
    b^{\star}_u = \arg\max_{b \in \mathcal{B}} \; \gamma_{u,b},
\label{eq:sinr_assoc}
\end{equation}
where $b^{\star}_u$ denotes the serving BS for user $u$.

\emph{ii)} The BS must have a sufficient number of resource blocks (RBs) available for the user to be connected:
\begin{equation}
    \zeta_b^{\mathrm{used}} + \zeta_{\mathrm{per\text{-}user}} \leq \zeta_b^{\mathrm{T}},
\label{eq:rb_constraint}
\end{equation}
where $\zeta_b^{\mathrm{used}}$, $ \zeta_{\mathrm{per\text{-}user}}$, and $\zeta_b^{\mathrm{T}}$ represent the number of used RBs, the RBs per user, and the total RBs of the BSs, respectively.

\emph{iii)} The signal strength of users (i.e., received power) must be higher than the receiver reference sensitivity, such that 
\begin{equation}
    \iota_{u,b} \geq \iota_{\mathrm{ref}},
\label{eq:sensitivity_constraint}
\end{equation}
where $\iota_{u,b}$ and $\iota_{\mathrm{ref}}$ denote the received signal strength from BS $b$ to user $u$, and the receiver reference sensitivity, respectively. The received signal strength $\iota_{u,b}$ is expressed as

\begingroup\setlength{\arraycolsep}{2pt}\renewcommand{\arraystretch}{1.05}
\begin{equation}
\label{eq:Iub_cases}
\iota_{u,b} =
\begin{cases}
\displaystyle
P_{\mathrm{T},i} - L_{\mathrm{LoS}},  & \text{TN (LoS)}, \\[1mm]
\displaystyle
P_{\mathrm{T},i} - L_{\mathrm{NLoS}}, &  \text{TN (NLoS)}, \\[1mm] 
P_{\mathrm{T},i} - L_{\mathrm{HAPS-SMBS}}, &  \text{NTN (HAPS-SMBS)}. \\[1mm]

\end{cases}
\end{equation}
\endgroup

\section{Problem Formulation}
\label{sec:problem formulation}

This section formulates SBS switching problem as an optimization problem that minimizes total network power consumption under load and sensitivity constraints, while reducing the number of unconnected users. The state vector $\Delta = [\delta_1, \delta_2, \dots, \delta_\Gamma]^\intercal$ represents the ON/OFF states of SBSs. Based on the conventional energy-focused model (EFM)~\cite{9}, two multi-objective optimization models are developed, namely WSM and $\epsilon$CM.
\subsection{Conventional Energy-Focused Cell Switching Problem Formulation (EFM)}
The cell switching problem has been predominantly modeled as the minimization of network power consumption~\cite{10}, i.e., 
\begin{subequations} 
\setcounter{equation}{0}

\begin{align}
\min_{\Delta} \quad & P(\Delta)\tag{10}\\[0.5em]
\text{s.t.} \quad
& \lambda_{\mathrm{H},t} \leq 1, \label{P1:a}\\
& \lambda_{\mathrm{M},t} \leq 1, \label{P1:b}\\
& \delta_{i,t} \in \{0,1\}, \quad i = 1,\ldots,s, \label{P1:c}
\end{align}
\label{P1}
\end{subequations}

\vspace{-0.5em}
\noindent
where $\lambda_{\mathrm{H},t}$ and $\lambda_{\mathrm{M},t}$ denote the load factors of the HAPS-SMBS and MBS, respectively. The total network power consumption is defined as
\begin{equation}
\begin{split}
    P(\Delta) = (P_{\mathrm{O,H}} + \eta_\mathrm{H} \lambda_\mathrm{{H},t} P_{\mathrm{T,H}}) 
           + (P_{\mathrm{O,M}} + \eta_\mathrm{M} \lambda_\mathrm{{M},t} P_{\mathrm{T,M}})\\\quad + \sum_{i=1}^{s} 
  \Big[ (P_{\mathrm{O},i} + \eta_i \lambda_{i,t} P_{\mathrm{T},i}) \, \delta_{i,t} 
        + P_{S,i}(1 - \delta_{i,t}) \Big].
\label{eq:total_power}
\end{split}    
\end{equation}
\subsection{Multi-Objective Optimization Problem Formulation}
The conventional energy-focused cell switching model in \eqref{P1} overlooks key performance metrics such as unconnected and dissatisfied users, which tend to increase after cell switching due to changes in propagation conditions. Since switching BSs can reduce SINR and data rates, incorporating these factors as objectives or constraints enables a balance between power consumption and achievable data rate. Accordingly, we define another optimization problem to minimize the number of unconnected users during cell switching as
\begin{subequations}\label{P2}
\noindent\textbf{} 
\label{P2}
\begin{align}
\setcounter{equation}{0}
{\min_{\Delta}} \quad & U_t(\Delta) \tag{12} \\[0.5em]
\text{s.t.} \quad
& \text{(10a) -- (10c)}, \nonumber \\
& \iota_{u,b} \geq  \iota_{\text{ref}}, 
\quad \forall u \in \mathcal{U},\; \forall b \in \mathcal{B}.  \label{P2:a}
\end{align}
\end{subequations}

\vspace{-0.5em}
\noindent
The feasibility indicator $f_{u,b}(\Delta)$ ensures that users can only associate with active BSs that satisfy receiver sensitivity and available RB constraints, and can be expressed as 
\begin{equation}
f_{u,b}(\Delta) = 
\delta_{i,t} \,
\mathbbm{1}\{\iota_{u,b} \geq \iota_{\text{ref}}\} \,
\mathbbm{1}\{\zeta_b^{\text{used}}(\Delta) + \zeta^{\text{per}} \leq \zeta_b^{\text{tot}} \delta_{i,t}\}.
\label{eq:fub}
\end{equation}
If $f_{u,b}(\Delta) = 0$, $\forall b$, the user is regarded as unconnected. The number of unconnected users is calculated as the sum of the non-connection indicators $z_u(\Delta)$, such that

\begin{equation}
z_u(\Delta) = 
\mathbbm{1}\left\{
\sum_{b \in \mathcal{B}} f_{u,b}(\Delta) > 0
\right\},
\label{eq:su}
\end{equation}
where $z_u(\Delta) = 1$ indicates that user $u$ is connected, and $z_u(\Delta) = 0$ indicates that user $u$ is unconnected. The total number of unconnected users is defined as

\begin{equation}
U(\Delta) = \sum_{u \in \mathcal{U}} \big( 1 - z_u(\Delta) \big).
\label{eq:U_obj}
\end{equation}

The other optimization problem is to minimize the number of dissatisfied users when cell switching is applied, which can be modeled as


\begin{subequations}
\noindent
\label{P3}
\begin{align}
\min_{\Delta} \quad & D(\Delta)  \tag{16}\\[0.5em]
\text{s.t.} \quad 
& \text{(10a) -- (10c)}, \nonumber  \\
& \text{(12a)} \nonumber.
\end{align}
\end{subequations}
A dissatisfied user is defined as one whose data rate after cell switching ($R_\mathrm{a}$) is lower than before cell switching ($R_\mathrm{b}$), i.e., $R_\mathrm{a} < R_\mathrm{b}$.

\subsubsection{Weighted Sum Method (WSM)}

In this method, the multi-objective optimization method, constructed by combining three individual optimization problems given in \eqref{P1}, \eqref{P2}, and \eqref{P3}, is expressed as

\noindent

\begin{equation}
\min_{\Delta} \;
\alpha \frac{P(\Delta)}{P_{\max}}
+ \beta \frac{U(\Delta)}{|\mathcal{U}|}
+ \upsilon \frac{D(\Delta)}{|\mathcal{U}|}
\label{M1}
\end{equation}
\[
\text{s.t.} \quad (10c), \nonumber  
\]
where $\alpha$, $\beta$, and $\upsilon$ represent the weights of the power consumption, the number of unconnected users, and the number of dissatisfied users optimization objectives, respectively, and each coefficient takes a value in the range $\left[ 0,1 \right]$. In addition, $P_{\max}$ and $\mathcal{U}$ denote the normalization parameters for the maximum network power consumption and the total number of users, respectively. Moreover, WSM employs a flexible and adaptive weighting mechanism that can be customized to different network requirements, enabling dynamic prioritization of objectives according to specific needs.  

\subsubsection{$\epsilon$-Constraint-Inspired Method ($\epsilon$CM)}

This method can be formulated by converting the optimization problems in \eqref{P2} and \eqref{P3} into constraints for the problem in \eqref{P1}. Hence, the optimization problem in \eqref{P1} is reformulated as

\begin{subequations}
\noindent 
\label{M2}
\begin{align}
\min_{\Delta} \quad & P(\Delta)  \tag{18}\\[0.5em]
\text{s.t.} \quad 
& \text{(10a) -- (10c)}, \nonumber   \\
& R_\mathrm{a} \geq R_\mathrm{b},  \label{M2:a} \\
& U_\mathrm{a} \leq U_\mathrm{b}, \label{M2:b} 
\end{align}
\noindent
\end{subequations}
where $U_\mathrm{a}$ and $U_\mathrm{b}$ are the numbers of unconnected users after and before cell switching, respectively.

Users initially connect to the BS with the highest SINR; after cell switching, however, the optimal BS for some users may be switched off, forcing them to associate with  more distant BSs. This increases path loss and may reduce the received power. Consequently, maintaining user connectivity becomes more challenging, directly affecting the objective in~\eqref{P2}, which aims to minimize the number of unconnected users. The resulting BS assignment issue after cell switching can be formulated as
\begin{equation}
b_u^{(\mathrm{bef})} = j \;\longrightarrow\; b_u^{(\mathrm{aft})} \in \{\text{MBS, HAPS-SMBS}\},
\end{equation}
where $b_u^{(\mathrm{bef})}$ and $b_u^{(\mathrm{aft})}$ represent the BS serving user $u$ before and after cell switching. $\Delta L_u$, the additional path loss due to switching, is given by
\begin{equation}
\Delta L_u \triangleq L_{u,b_u^{(\mathrm{aft})}} - L_{u,b_u^{(\mathrm{bef})}} \geq 0,
\end{equation}
which is generally a positive value because of the increased distance between the user and the serving BS. $\iota_{u,b}^{(\mathrm{aft})}$, is the received signal strength after cell switching, is given by
\begin{equation}
\iota_{u,b}^{(\mathrm{aft})} = \iota_{u,b}^{(\mathrm{bef})} - \Delta L_u,
\end{equation}
where  $\iota_{u,b}^{(\mathrm{bef})}$ denote the received signal strength before cell switching. If the $\iota_{u,b}^{(\mathrm{aft})}$  falls below $\iota_{\mathrm{ref}}$ following cell switching, the user fails to establish a connection and is classified as an unconnected user. Consequently, the objective of minimizing the number of unconnected users in \eqref{P2} is not satisfied. In this context, our model addresses this issue using \eqref{M1} and \eqref{M2}.

Another problem that arises after cell switching is the reduction in data rate caused by a decline in SINR. If the data rate after cell switching is lower than that before cell switching, the user is considered dissatisfied. In this case, the problem in \eqref{P3} emerges due to of the increase in the number of dissatisfied users.

\section{Performance Evaluation}
\label{sec:Performance Evaluation} 

To compare network performance, the achievable data rate is analyzed for BEL values from 0 to 30 dB~\cite{ITU_R_P2346_3}. Atmospheric loss is considered constant due to its negligible variation, placing the focus on BEL. Three user types are defined to capture propagation-aware performance: high-loss indoor, low-loss indoor, and outdoor users. High-loss indoor users experience additional attenuation beyond BEL, low-loss indoor users experience reduced extra loss as defined in~\eqref{loss}, and outdoor users experience no BEL. Each user is allocated one RB, resulting in a constant traffic demand per user during each simulation interval and generating the overall network traffic. These propagation and traffic models influence the optimization by affecting the achievable data rate and consequently the cell switching decisions.

\begin{figure}[H]
    \centering
    \includegraphics[width=0.9\linewidth]{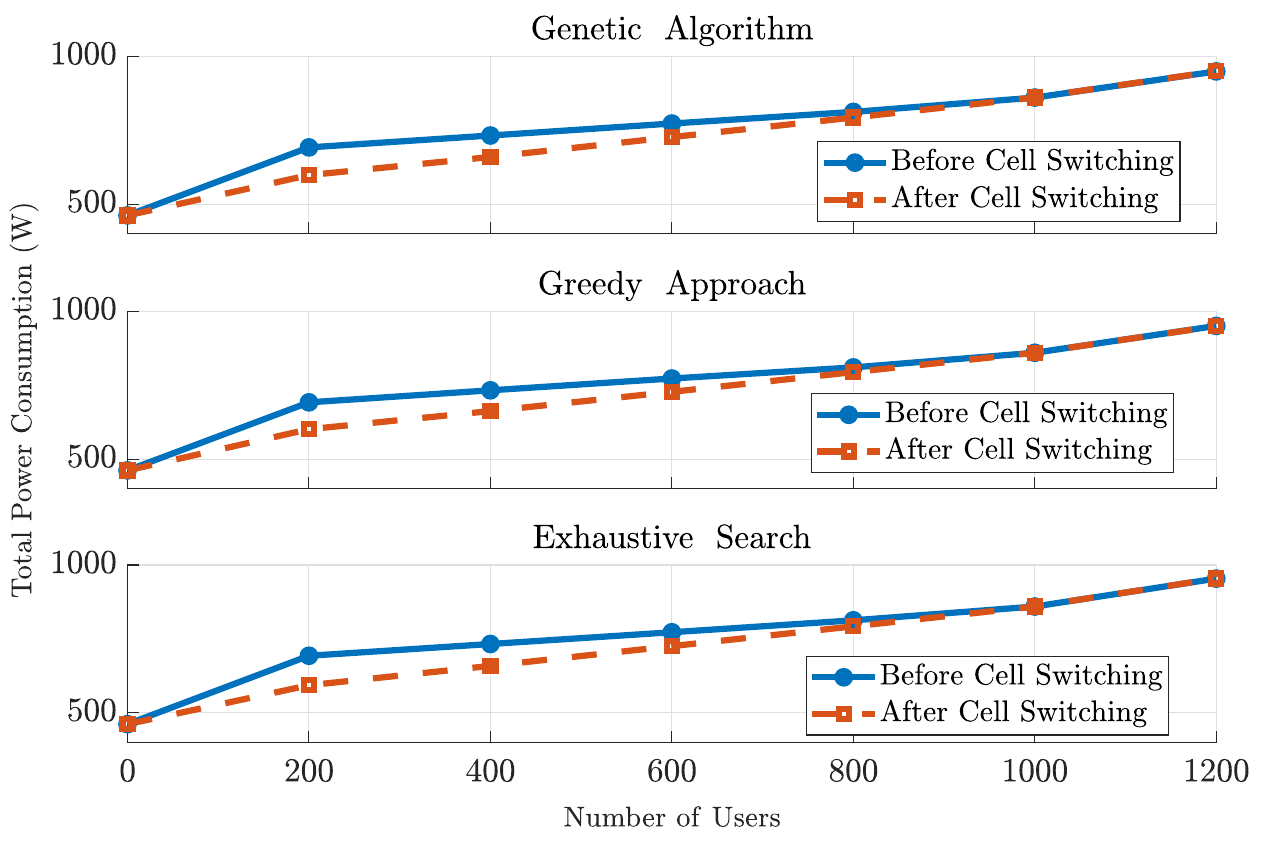}
    \caption{Total network power consumption of the $\epsilon$CM method with three different methodologies (i.e., genetic algorithm, greedy approach, and exhaustive search).}
    \label{comparison}
\end{figure}

As shown in Fig.~\ref{comparison}, three cell switching algorithms are evaluated in the $\epsilon$CM framework: genetic algorithm, greedy approach and exhaustive search. A genetic algorithm is employed as a metaheuristic approach, where each chromosome represents an SBS ON/OFF pattern, enabling the identification of near-optimal solutions while satisfying the $\epsilon$-constraints. In addition, a greedy algorithm is employed that iteratively switches off SBSs with the least marginal impact on the objective, achieving performance close to that of exhaustive search with significantly lower complexity. Exhaustive search tests all ON/OFF SBS combinations to find the optimal configuration under $\epsilon$-constraints, but its complexity grows exponentially (O($2^{\Gamma}$)) with the number of SBSs since all possible combinations must be evaluated, limiting its use in dense scenarios. Despite its higher complexity, exhaustive search is selected as the primary algorithm due to the small network size, while greedy and genetic methods are employed only for comparison.

\begin{figure*}[t]
\centering
\begin{subfigure}{0.32\textwidth}
  \centering
  \includegraphics[width=0.90\linewidth]{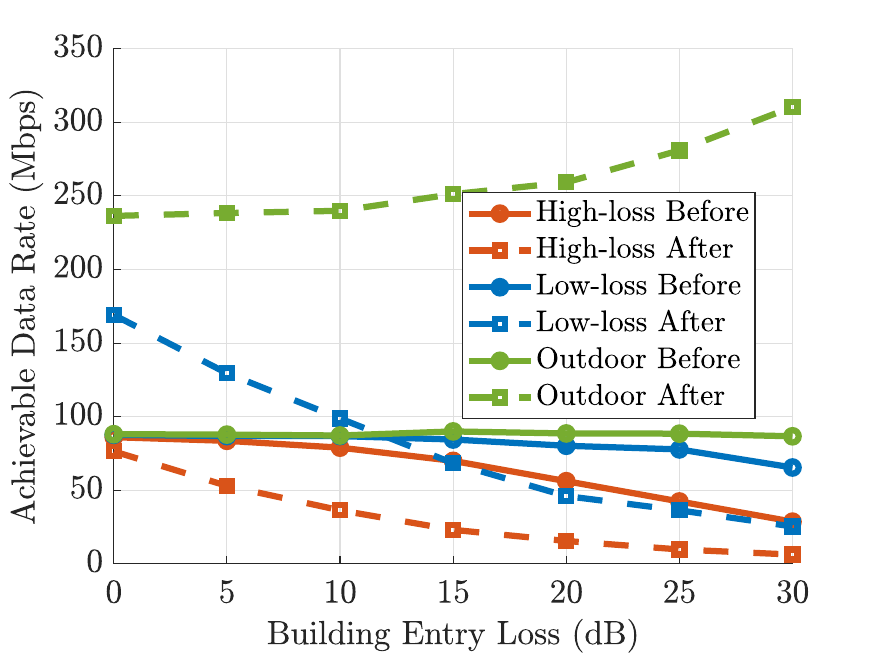}
  \caption{EFM}
  \label{T3a}
\end{subfigure}\hfill
\begin{subfigure}{0.32\textwidth}
  \centering
  \includegraphics[width=0.90\linewidth]{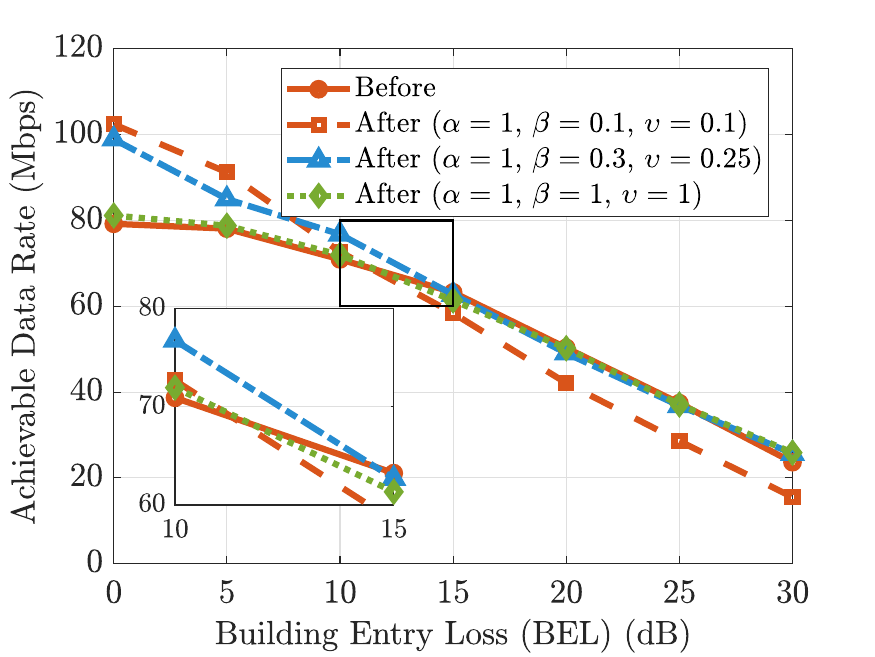}
  \caption{WSM}
  \label{T3b}
\end{subfigure}\hfill
\begin{subfigure}{0.32\textwidth}
  \centering
  \includegraphics[width=0.90\linewidth]{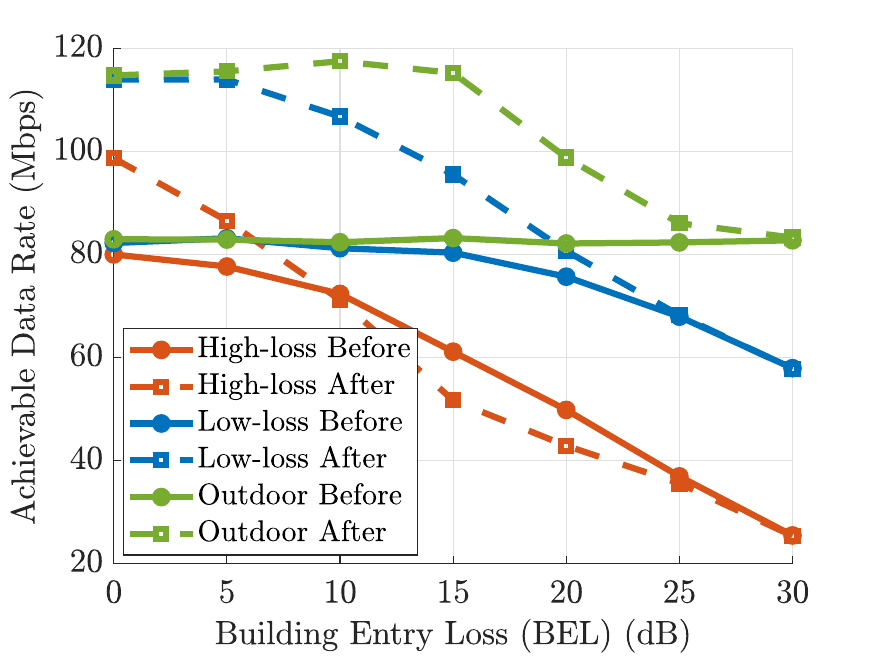}
  \caption{$\epsilon$CM}
  \label{T3c}
\end{subfigure}
\caption{Achievable data rate of EFM, WSM, and $\epsilon$CM methods under different user densities.}
\label{fig:data}
\end{figure*}

\begin{figure*}[b]
\centering
\begin{subfigure}{0.32\textwidth}
  \centering
  \includegraphics[width=0.90\linewidth]{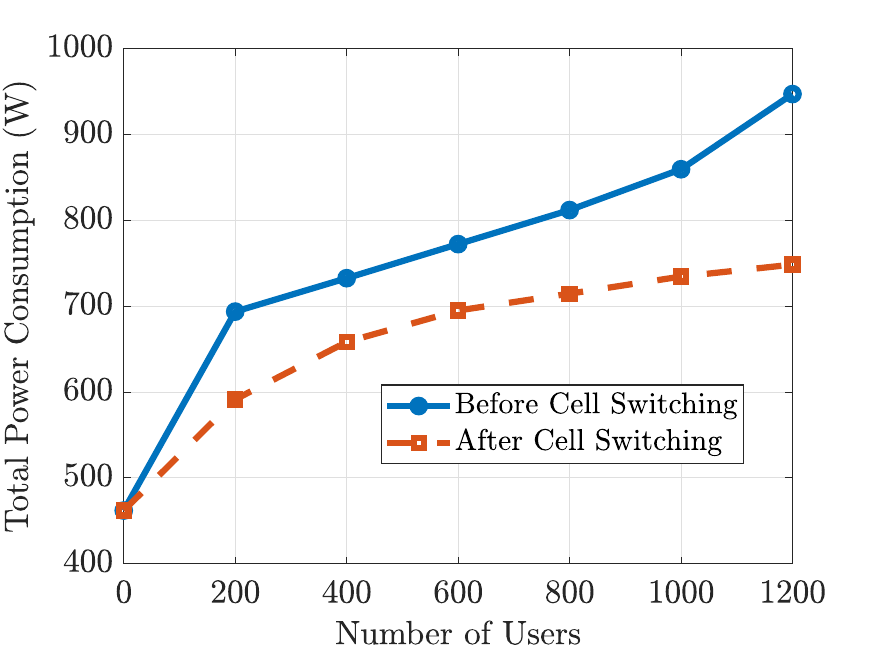}
  \caption{EFM}
  \label{P3a}
\end{subfigure}\hfill
\begin{subfigure}{0.32\textwidth}
  \centering
  \includegraphics[width=0.90\linewidth]{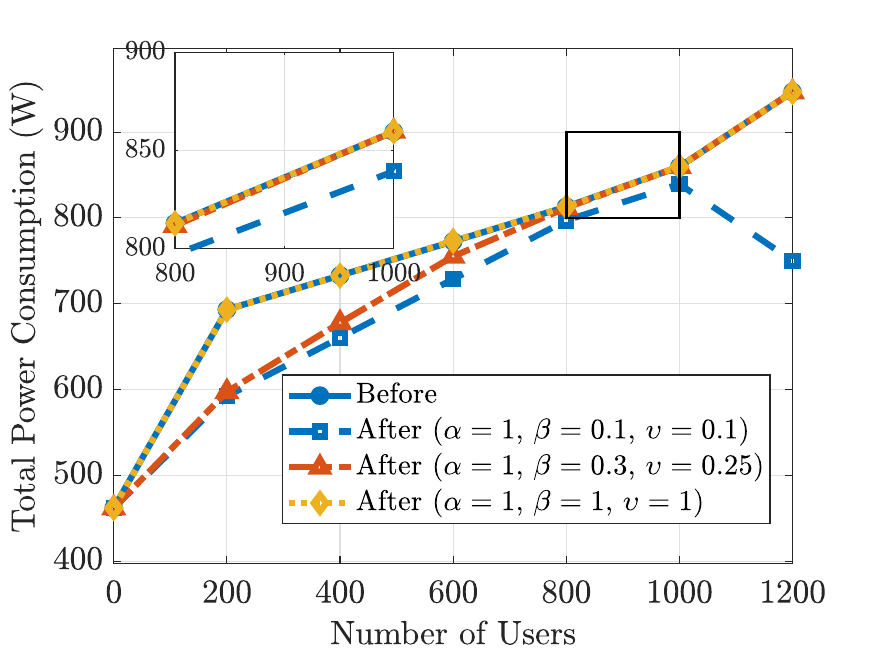}
  \caption{WSM}
  \label{P3b}
\end{subfigure}\hfill
\begin{subfigure}{0.32\textwidth}
  \centering
  \includegraphics[width=0.90\linewidth]{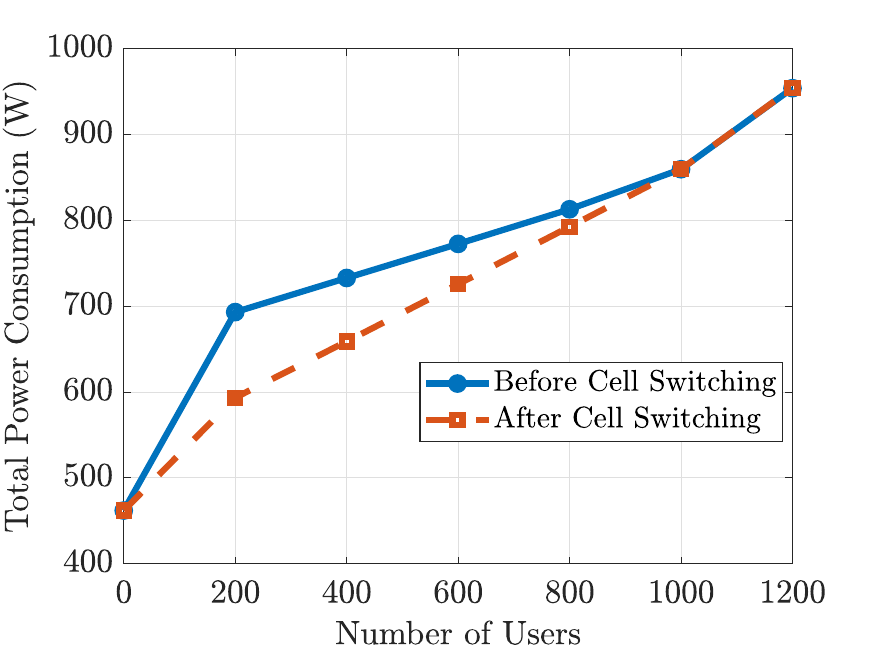}
  \caption{$\epsilon$CM}
  \label{P3c}
\end{subfigure}
\caption{Total power consumption of EFM, WSM, and $\epsilon$CM methods under different user densities.}
\label{fig:energy}
\end{figure*}

As shown in Fig. \ref{T3a}, after cell switching using EFM, high-loss indoor users experience decreasing achievable data rates as BEL increases due to their high sensitivity to propagation loss. The results are obtained using exhaustive search as a reference method to ensure a fair comparison of the proposed formulations. In contrast, low-loss indoor users observe an increase in achievable rates up to a certain BEL value. This improvement occurs because switching off specific BSs reduces the interference power $P_{u,k}$ in the denominator of \eqref{SINR}, thereby increasing SINR and the achievable rate. However, beyond approximately 15 dB of BEL, the interference reduction can no longer compensate for the increased propagation loss, leading to a decline in achievable rate. Outdoor users benefit from reduced interference and therefore achieve higher data rates. When BEL reaches around 20 dB, the reduced connectivity of indoor users allows outdoor users to access more MBS or HAPS-SMBS RBs, resulting in a noticeable increase in their achievable rates after cell switching.

Fig. \ref{T3b} presents the achievable data rate across different BEL values under the WSM. Since user types are shown elsewhere and WSM uses weight factors, the results focus on high-loss indoor users to highlight QoS-dominant, balanced, and power-dominant scenarios. The coefficients are selected to illustrate three distinct operating regimes of the WSM, and their specific values are determined empirically. In the QoS-dominant case ($\alpha = 1$, $\beta = 1$, $\upsilon = 1$), the optimization equally penalizes unconnected and dissatisfied users, so most BSs remain active and the achievable rate remains unchanged before and after cell switching. In the balanced scenario ($\alpha = 1$, $\beta = 0.3$, $\upsilon = 0.25$), cell switching provides a gain in data rate up to 10 dB of BEL, while mitigating degradation caused by increasing BEL. In the power-dominant case ($\alpha = 1$, $\beta = 0.1$, $\upsilon = 0.1$), WSM prioritizes power minimization and switches off BSs without considering unconnected or dissatisfied users, causing the achievable rate to decrease as BEL increases.

Fig. \ref{T3c} illustrates the achievable data rate across different BEL values in $\epsilon$CM. For high-loss indoor users, cell switching provides a data rate gain up to approximately 10 dB of BEL, after which increases in BEL and BS deactivation to minimize power under user constraints lead to reduction in data rate. Low-loss indoor users experience increased data rates up to 25 dB of BEL; beyond this, their rates remain nearly unchanged as they are less affected by BEL. Outdoor users achieve higher data rates after cell switching across all BEL values due to reduced interference. However, at high BEL levels, $\epsilon$CM activates certain BSs to maintain indoor connectivity, which increases interference and slightly reduces outdoor data rates.

As shown in Fig.~\ref{P3a}, applying EFM results in lower power consumption across all user densities after cell switching. This occurs because EFM focuses solely on minimizing power consumption without accounting for dissatisfied or unconnected users, leading to BS deactivation whenever possible. However, once the number of users exceeds 1000, total power consumption stabilizes as the algorithm reaches the optimal number of active BSs.

Fig. \ref{P3b} shows three WSM scenarios: QoS-dominant, balanced, and power-dominant. In the QoS-dominant case ($\alpha=1$, $\beta=1$, $\upsilon=1$), the optimization prioritizes user connectivity, keeping BSs active and preventing power savings. In the balanced scenario (i.e., $\alpha = 1$, $\beta = 0.3$, and $\upsilon = 0.25$), power consumption remains lower after cell switching until the number of users reaches 800. In the power-dominant scenario  (i.e., $\alpha = 1$, $\beta = 0.1$, and $\upsilon = 0.1$), power consumption is reduced across all user densities.

Fig.~\ref{P3c} shows that in $\epsilon$CM, power consumption decreases after cell switching for up to 400 users, as fewer users allow more BSs to be switched off. Between 400 and 800 users, the optimization prioritizes minimizing dissatisfied and unconnected users, leading to gradual BS reactivation and increased power consumption. Beyond 800 users, most BSs remain active, and total power approaches the pre-switching level.

\begin{figure}
    \centering
    \includegraphics[width=0.8\linewidth]{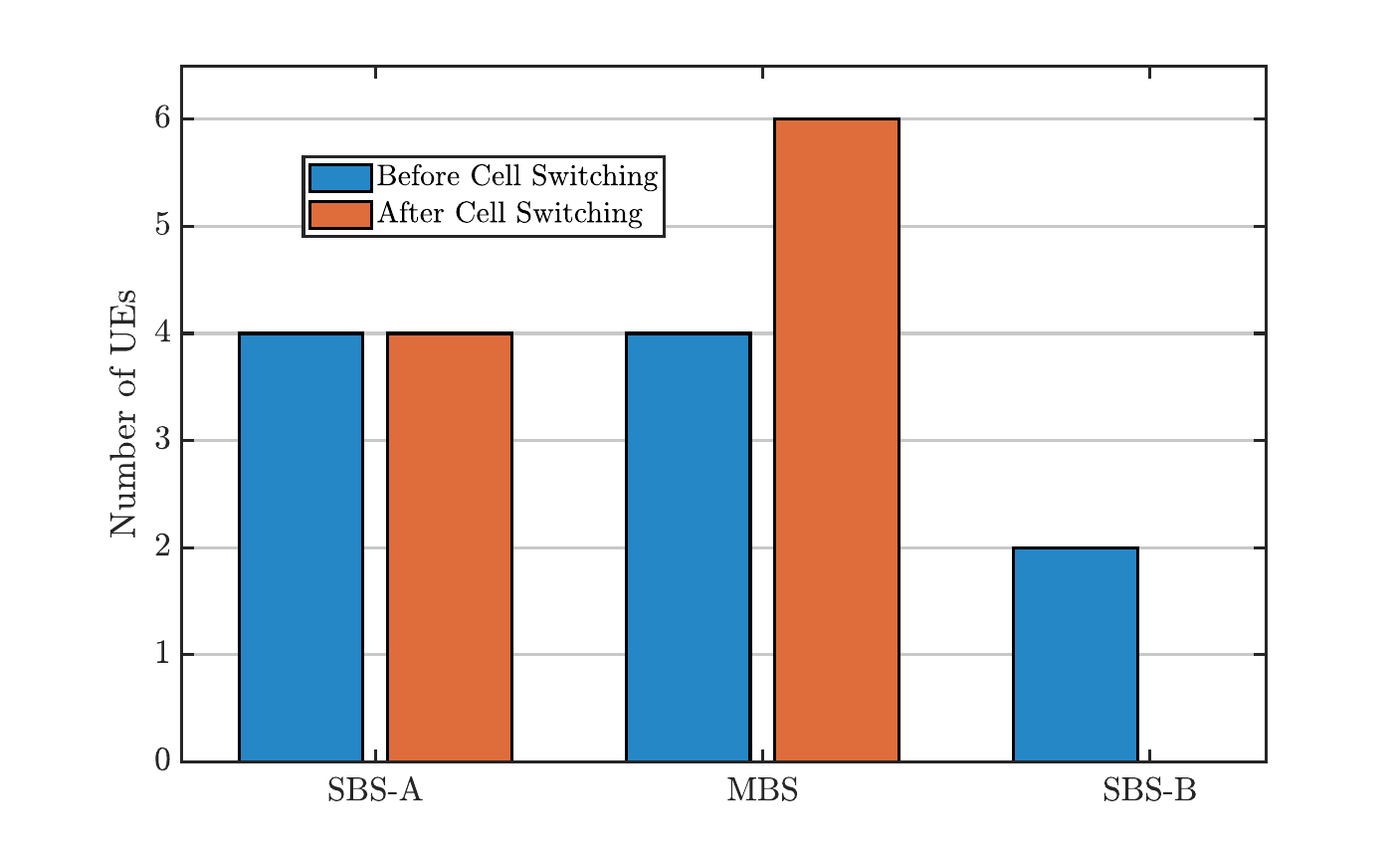}
    \caption{Sionna-OAI cell switching framework.}
    \label{OAI}
\end{figure}

Fig.~\ref{OAI} presents a simplified example with 3 BSs and 10 UEs to illustrate the cell switching framework implemented in the Sionna–OAI environment, where simplification reduces computational complexity. In this framework, the $\epsilon$CM method monitors instantaneous BS loads after UE association to make switching decisions. Initially, UEs are distributed among SBS-A, MBS, and SBS-B according to the association policy. After switching, the least-loaded SBS2 is switched off, and its UEs are offloaded to the MBS. This example illustrates the $\epsilon$CM framework in Sionna–OAI, where OAI provides association and Sionna evaluates SINR feasibility.

\section{Conclusion}
\label{sec:Conclusion}

The conventional cell switching problem focused only on energy efficiency, overlooking connectivity, data rate, and environmental factors such as BEL for indoor users and atmospheric losses for NTN users. In this study, the conventional problem was reformulated for a HAPS-assisted network as a multi-objective optimization framework that considers unconnected and dissatisfied users through the WSM and $\epsilon$CM methods. The proposed approach was validated through system-level simulations and Sionna–OAI emulation. The results showed that WSM reduced or eliminated data rate degradation, while $\epsilon$CM minimized rate loss and power consumption. Overall, the proposed approaches improved sustainability without compromising connectivity or data rate by balancing energy efficiency and user performance.

\section*{Acknowledgment}
Mehmet Eren Uluçınar was supported by the Scientific and Technological Research Council of Türkiye (TÜBİTAK). This work was also supported by TÜBİTAK Informatics and Information Security Research Center (BİLGEM) under the HAPS-Com Project.

\bibliographystyle{IEEEtran} 
\bibliography{references}

\end{document}